\documentclass[12pt,a4paper]{article}

\newcommand{\be}{\begin{equation}}
\newcommand{\ee}{\end{equation}}

\let\abs=\envert

\usepackage{amssymb,amsmath}

\begin{document}
\title{\bf From 2-Dimensional Surfaces to Cosmological Solutions}

\author{A. Szereszewski and J. Tafel}
\date{}
\maketitle

\noindent
Institute of Theoretical Physics, University of Warsaw,
Ho\.za 69, 00-681 Warsaw, Poland, email: tafel@fuw.edu.pl

\bigskip

\bigskip

\noindent
{\bf Abstract}.\\ 
We construct perfect
 fluid metrics corresponding 
to spacelike surfaces invariant under a 1-dimensional group of isometries in 
3-dimensional Minkowski space.
Under additional assumptions we obtain new cosmological solutions of Bianchi type
 $II$, $VI_0$ and $VII_0$.
The solutions depend on an arbitrary function of time, which can be specified
 in order to satisfy an equation of state.

\noindent



\bigskip

\section{Introduction}
Perfect fluid solutions of the Einstein equations with a 2-dimensional group of isometries play an important role in nonhomogeneous cosmology \cite{akr}. Usually they are obtained under additional assumptions as separability of the metric coefficients, degeneracy of the Weyl tensor or existence of conformal symmetries (see \cite{S} and references therein). Some of these solutions are free of singularities \cite{s} \cite{rs} \cite{ms}. 

In \cite{ST} we presented a geometric construction of $G_2$  perfect
fluid metrics for which the Einstein equations reduce to a single differential equation. The metrics  
 are related
to 2-dimensional surfaces in 3-dimensional nonphysical Minkowski space $M^{2,1}$. In this paper we apply this method to 
surfaces which are invariant under a 1-dimensional group of isometries of $M^{2,1}$.  Under 
additional assumptions the corresponding metrics admit 3-dimensional 
groups of isometries. In this way we obtain tilted or nontilted cosmological solutions of Bianchi type $II$, $VI_0$ or $VII_0$. They are given up to  the differential equation. We solve this equation in quadratures in the case of vanishing tilt. The resulting metrics depend on a free function. Hence, a barotropic equation of state can be imposed, however this probably requires numerical calculations. Up to knowledge of
the authors the presented solutions are new \cite{S}.

For a completeness we describe below the basic construction of \cite{ST}.
Let  $M^{2,1}$ be the 3-dimensional Minkowski space  with coordinates $x^i,\ i=1,2,3$, and the metric 
$\eta_{ij}=$diag(1,1,-1). Consider a spacelike surface $\Sigma\subset M^{2,1}$ with the negative Gauss curvature $K$
\be
K<0\ .\label{K}
\ee
 The internal coordinates of $\Sigma$ are denoted by $x^A,\ A=0,1$. Let 
$\vec{n}=(n_i)$ be the normal vector of $\Sigma$ such that 
 \begin{align}
    \vec{n}^2=-1 \, , \, n_3<0 \, ,        \label{n2}
 \end{align}
where 
$\vec{n}^2=n_1^2+n_2^2-n_3^2$. We relate with $\Sigma$ the following  4-dimensional spacetime metric 
\begin{equation}
    g=\gamma e^{\psi}g'_{AB}dx^A dx^B+\rho\,n_{ab}dx^a dx^b,   \label{metric}
 \end{equation}
where $\gamma=\pm 1$, $g'_{AB}$ is a fixed metric proportional to the second fundamental form  of $\Sigma$, functions $x^a,\ a=2,3$, together with $x^A$
compose a 4-dimensional system of coordinates and 
\begin{equation}
   \rho=|K|^{-1/2}, \label{K1}
 \end{equation}
\begin{equation}
           n_{ab}  =\begin{pmatrix}
                  n_3 +n_1   &  n_2 \\
                     n_2     &  n_3-n_1 
               \end{pmatrix} \, .          \label{n_ab}
 \end{equation}
Components of $g$ are independent of the coordinates $x^a$. Hence, $\partial_a$ are the Killing vector fields. 

Consider the following  equation
\begin{equation}  
   \det (\tilde R_{AB})=0\, ,   \label{detR=0}
 \end{equation}
where 
\begin{equation}
     \tilde{R}_{AB}=R_{AB}+\frac{\Box \rho}{2\rho}g_{AB} \, ,    \label{tildeRAB}
 \end{equation}
\begin{equation}
     R_{AB}=\frac{1}{2}R^{(2)}g_{AB}-\frac{1}{\rho}\rho_{\vert AB}+\frac{1}{2\rho^2}\rho_{,A}
          \rho_{,B}-\frac{1}{2}\vec{n}_{,A}\vec{n}_{,B}\ . \label{eqnRAB}
\end{equation}
Here the  symbol $_{\vert A}$ denotes the covariant derivative with respect to the metric 
$g_{AB}=\gamma e^{\psi}g'_{AB}$, $\Box \rho=\rho^{\vert A}_{\ A}$ and $R^{(2)}$ is the scalar curvature of $g_{AB}$. Given the surface $\Sigma$ equation (\ref{detR=0}) yields a complicated second order equation for $\psi$. If it is satisfied and 
\begin{equation}
\tilde R^A_{\ A}\tilde R_{00}>0\label{cond}
\end{equation}
then metric (\ref{metric}) is a perfect fluid  solution of the Einstein equations. 

In this approach the energy density $\epsilon$ and the pressure $p$ of the fluid are given by the following expressions (we omit the constant $8\pi G/c^4$)
 \begin{align}
  \epsilon&=\frac{1}{2}R^A\,_A+\rho^{-1}\Box \rho\ ,   \label{ee}\\
  p&=\frac{1}{2}R^A\,_A \, .                                           \label{p}
\end{align} 
If $\Box \rho=0$ then $p=\epsilon$. This case is described by a close analogue of the Ernst equation, which admits various solution generating methods \cite{S}. For this reason we will  focus here on the case $p\neq \epsilon$.

In order to satisfy the condition $p<\epsilon$ one should require
\begin{equation}
\gamma\Box'\rho>0\ ,\label{gamma}
\end{equation}
where $\Box'$ denotes the wave operator related to the metric $g'_{AB}$.
%\be
%\Box'\rho=\frac{1}{\sqrt{|g'|}}(\sqrt{|g'|}g'^{AB}\rho_{,A})_{,B}\ , \ |g'|=det(g'_{AB})\ .
%\ee
If one assumes 
\be
   \tilde{R}^A\,_A >0         \label{cond3}
 \ee
in addition to (\ref{cond}) and (\ref{gamma})
 then the energy dominant condition $\epsilon>|p|$
is satisfied.

In this paper we apply the above approach to surfaces admitting a continuous symmetry. 
In section 2 we classify surfaces $\Sigma$ invariant under a 1-dimensional group of isometries of $M^{2,1}$.  In sections 3-5 we calculate the corresponding metric tensors and we obtain solutions of equation (6) giving rise to homogeneous cosmological solutions of the Einstein equations.

\section{Invariant Surfaces}

Under  translations and Lorentz  transformations of coordinates $x^i$ of $M^{2,1}$ the induced metric of $\Sigma$, $g_I=dx_idx^i$, and its second fundamental form,
$g_{II}=dn_i dx^i$,  remain
unchanged. The same is true for metric (\ref{metric}) provided $\psi$ is a scalar function and coordinates
$x^a$ undergo a linear spinor transformation corresponding to the Lorentz one.

Now, assume that a surface $\Sigma$ is invariant under an active Poincar\'e transformation $P$ of 
$M^{2,1}$. Then the forms $g_I$ and $g_{II}$ are preserved. The same refers  to the metric (\ref{metric}) provided $\psi$ is 
invariant under $P$ and $x^a$ undergo the corresponding spinor transfromation. Thus, the symmetry of 
$\Sigma$ and $\psi$ induces a symmetry of the 4-dimensional spacetime metric $g$. Note that the 
invariance of $\psi$ is compatible with equation (\ref{detR=0}).

In what follows we assume that $\Sigma$ is invariant under a 1-dimensional group of isometries of 
$M^{2,1}$ corresponding to a Killing vector $k$ (we will call such $\Sigma$ an invariant surface). If $\psi$ is
also invariant  the metric $g$ acquires a new continuous symmetry  in addition to the 
symmetries $\partial_a$.

 A general Killing field $k$ in $M^{2,1}$ has the form
\begin{equation}
k=(\epsilon^{ijl}A_ix_j+B^l)\partial_l\ ,\label{k}
\end{equation}
where $\vec{A}$, $\vec{B}$ are constant vectors, $\epsilon^{ijl}$ is the completely antisymmetric tensor
and indices $i,j$ are raised or lowered by means of the 3-dimensional Minkowski metric $\eta_{ij}$. By a
Lorentz rotation the vector $\vec{A}$ can be put into one of three canonical forms depending whether it is timelike (T),
spacelike (S)  or null (N). Then one can shift coordinates $x^i$ in order to simplify vector
$\vec{B}$. In this way one obtains the following canonical forms of the Killing fields with a  
nonvanishing rotational part (only such symmetries are admitted by surfaces with a nonvanishing curvature $K$)
\begin{flalign}
&\text{T}: &k&=-x_2\partial_1+x_1\partial_2+b\partial_3\label{k1}\, ,\\
&\text{S}: &k&=-x_3\partial_2+x_2\partial_3+b\partial_1\label{k2}\, ,\\
&\text{N}: &k&=(x_1-x_3)\partial_2+x_2(\partial_3-\partial_1)+b(\partial_1+\partial_3)\ ,
\label{k3}
\end{flalign}
where $b$ is a constant.

The Killing field $k$ has to be tangent to the surface $\Sigma$ if the latter is invariant under the 
1-parameter group of transformations generated by $k$. Hence, the surface can be defined by an equation 
$F(I_1,I_2)=0$, where $I_1,\ I_2$ are two independent solutions of the equation $k^iI_{,i}=0$ in 
$M^{2,1}$. For fields (\ref{k1})-(\ref{k3}) invariant surfaces with nonzero curvature are 
given, in some coordinates $x^i$, by one of the following sets of relations:
 \begin{flalign}
&\text{T:}   &x&^1=\tau \cos{\varphi} ,\quad  &x&^2=\tau \sin{\varphi} , \quad
&x&^3=a(\tau)+b\varphi ,       &\tau\geq 0 ,       \label{s1} \\
&\text{S1: }  &x&^1=a(\tau)+b\varphi ,\quad  &x&^2=\tau \cosh{\varphi} ,\quad  
                &x&^3=\tau \sinh{\varphi} ,        &\tau\geq 0 ,       \label{s2}\\ 
&\text{S2: }  &x&^1=a(\tau)+b\varphi ,\quad  &x&^2=\tau \sinh{\varphi} ,\quad 
                &x&^3=\tau \cosh{\varphi} ,       &\tau\geq 0 ,        \label{s3}
 \end{flalign}
\begin{flalign}
&\text{N1}: \frac{v^2}{2b}-2x_2=a\Big(u+\frac{vx_2}{b}-\frac{v^3}{6b^2}\Big)
, &x^1&=\frac{1}{2}(v+u), &x^3&=\frac{1}{2}(v-u)      \label{s6}\\
&\text{N2}: u+\frac{x_2^{\ 2}}{v}=a(v), &x^1&=\frac{1}{2}(v+u), &x^3&=\frac{1}{2}
(v-u) \label{s5}\\
&\text{N3}:   u+\frac{vx_2}{b}-\frac{v^3}{6b^2}=\text{const}, &x^1&=\frac{1}{2}(v+u), &x^3&=\frac{1}{2}(v-u),  \label{s4}
\end{flalign}
where $a$ is an arbitrary function of one variable and $b\neq 0$ in the cases N1 and N3. 

In sections 3-5 we derive metrics corresponding to surfaces (\ref{s1})-(\ref{s4}).

\section{Metrics related to surfaces of type T}
Due to a residual freedom in choice of coordinates $x^i$  for surfaces (\ref{s1}) one can assume
\begin{equation}
b\ge 0\label{b}
\end{equation}
without a loss of generality.
 In order to guarantee the spacelike character of $\Sigma$ the function $a(\tau)$ should satisfy
 \begin{equation}
   \dot{a}^2<1-b^2\tau^{-2}\, ,\label{cond4}
 \end{equation} 
where the dot denotes the derivative with respect to $\tau$. 
The normal vector  satisfying condition (\ref{n2}) reads
 \begin{equation}
  \vec{n}=\frac{1}{\sqrt{\tau^2(1-\dot{a}^2)-b^2}}(\tau \dot{a}\cos\varphi-b \sin\varphi,
         \tau \dot{a}\sin\varphi+b \cos\varphi, -\tau)\ .
 \end{equation}
The second fundamental form is given by 
 \begin{align}
   g_{II}&=\frac{1}{\sqrt{\tau^2(1-\dot{a}^2)-b^2}}(\tau\ddot{a}d\tau^2-2b\, d\tau d\varphi
          +\tau^2\dot{a}d\varphi^2)\ .
 \end{align}
The Gauss curvature $K=\det{g_{II}}/\det{g_I}$ is negative provided
 \begin{equation}
   \dot{a}\ddot{a}<b^2\tau^{-3} \, . \label{a'a''}
 \end{equation}
Under conditions (\ref{cond4}) and (\ref{a'a''}) definition (\ref{K1}) yields
 \begin{align}
   \rho&=\frac{\tau^2(1-\dot{a}^2)-b^2}{\sqrt{b^2-\tau^3\dot{a}\ddot{a}}} \, .
 \end{align}
In order to describe the corresponding spacetime metric (\ref{metric}) it is convenient to use coordinates $\tau$, $x$, $y$, $z$, where $x^a=x,y$ and 
\begin{equation} 
z=\frac{1}{2}(\varphi-\sigma)\ ,\ \ \sigma (\tau)=b\int \frac{d\tau}{\tau^2\dot{a}}\ .\label{chi}
\end{equation}
In these coordinates the metric takes  the following form
 \begin{align}
 g=&\gamma e^{\psi}\Big[\big(\frac{b^2}{4\tau^4\dot{a}^2}-\frac{\ddot{a}}{4\tau \dot{a}}\big)d\tau^2-
     dz^2\Big]-\sqrt{\frac{\tau^2(1-\dot{a}^2)-b^2}{b^2-\tau^3\dot{a}\ddot{a}}}\big[(\tau-\tau\dot{a}
     \cos \sigma+b\sin \sigma)\nonumber\\
   &(\theta^2)^2-2(\tau\dot{a}\sin \sigma+b\cos \sigma)\theta^2 \theta^3+(\tau+\tau
     \dot{a}\cos \sigma-b\sin \sigma)(\theta^3)^2\big]\ ,\label{metric1} 
 \end{align}
where
 \begin{align}
     &\theta^2=\cos z dx+\sin z dy\ ,\quad \theta^3=-\sin z dx+\cos z dy\ .\label{theta}
 \end{align}
 The Einstein equations for metric (\ref{metric1}) reduce to equation (\ref{detR=0}) and condition (\ref{cond}). 

Under the assumption 
 \begin{equation}
   \gamma=1,\  \psi=\psi(\tau)      \label{psi_t}
 \end{equation}
metric (\ref{metric1}) is a cosmological solution of Bianchi type $VII_0$. In the generic case 
the fluid velocity is not orthogonal to the surfaces of homogeneity $\tau=$const (so called tilted solution). Since metric (\ref{metric1}) depends on two functions $a(\tau)$, $\psi (\tau)$ one can impose on $p$ and $\epsilon$ 
a barotropic equation of state
 \begin{equation}
   p=p(\epsilon)\ .      \label{eqn_st}
 \end{equation}
Equations (\ref{detR=0}), (\ref{eqn_st}) together with condition (\ref{cond}) and, if required, conditions (\ref{gamma}), (\ref{cond3})
compose a  system of equations for functions $a(\tau)$ and $\psi(\tau)$. 

When (\ref{psi_t}) is satisfied and   
 \begin{equation}
    b=0  \label{b=0}
 \end{equation} 
the fluid velocity is orthogonal to the surfaces $\tau$=const. Then   equation (\ref{detR=0}) reduces to 
 \begin{equation}
   \tilde{R}_{11}=0     \label{R11=0}
 \end{equation}
and inequality (\ref{cond}) becomes a part of the energy condition
 (\ref{cond3}).
Equation (\ref{R11=0}) yields 
 \begin{equation}
   \psi=\log \rho-\int \frac{(a-a_0)\ddot{a}}{\tau(1-\dot{a}^2)}d\tau \, ,
 \end{equation}
where  $a_0$ is a constant and 
 \begin{equation}
   \rho=(1-\dot{a}^2)\sqrt{-\frac{\tau}{\dot{a}\ddot{a}}}  \label{rho}
 \end{equation}
(note that $|\dot a|<1$, $\dot{a}\ddot{a}<0$ due to (\ref{cond4}) and 
(\ref{a'a''})). 
The corresponding  spacetime metric reads 
 \begin{equation}
  \begin{split}
   g=&\rho e^{\omega}\Big(-\frac{\ddot{a}}{4\tau\dot{a}}d\tau^2-dz^2\Big)-\rho\sqrt{\frac{1-\dot{a}}{1+\dot{a}}}(\cos z dx
     +\sin z dy)^2+\\
     &-\rho\sqrt{\frac{1+\dot{a}}{1-\dot{a}}}(-\sin z dx+\cos z dy)^2 \, ,         \label{sol1}
  \end{split}
 \end{equation}
where function $\omega(\tau)$ is given by
 \begin{align}
  \omega&=-\int \frac{(a-a_0)\ddot{a}}{\tau(1-\dot{a}^2)} d\tau\, . \label{omega}
 \end{align}
In this case expressions (\ref{ee}) and (\ref{p}) for the fluid energy and pressure yield
 \begin{align}
  \epsilon(\tau)&=\rho^{-2}e^{-\omega}\Big(\frac{3\rho'\,^2}{4\rho}+\frac{\lambda\rho'}{\rho}(a-a_0)-\frac{\lambda a'}{2}-\frac{\tau^2}{\rho}\Big) \, ,    \label{e1}\\
  p(\tau)&=\epsilon(\tau)-\frac{\rho''}{\rho^2}e^{-\omega} \, .              \label{p1}
 \end{align}
Here $\lambda=\text{sgn}(\dot{a})$ and we denote 
\begin{equation}
f'=\dot f \sqrt{\frac{4\tau|\dot{a}|}{|\ddot{a}|}}\label{f'}
\end{equation}
for any function $f(\tau)$.
 Expressions (\ref{e1}), 
(\ref{p1}) contain one arbitrary function $a(\tau)$, which can be fitted in order  to satisfy a barotropic equation
of state (\ref{eqn_st}) of the fluid. Due to the complicated structure of (\ref{e1}) and (\ref{p1}) this fitting probably requires numerical calculations.
On the other hand it is not difficult to find examples of $a(\tau)$ for which the energy dominant condition is satisfied. Such example was given in \cite{ST}. 
In this example the asymptotic behaviour of $p$ and $\epsilon$ is similar to that in the model of Demia\'nski and Grishschuk \cite{dg}.

\section{Metrics related to surfaces of type S}
In this section we consider metrics related to the invariant surfaces (\ref{s2}) and (\ref{s3}).  In both cases the second fundamental form is proportional to
 \begin{equation}
g'_{AB}dx^Adx^B=\frac{\ddot{a}}{4\tau\dot{a}}d\tau^2-\frac{b}{2\tau^2\dot{a}}d\tau d\varphi-\frac{1}{4}d\varphi^2 \, ,
 \end{equation}
however the corresponding spacetime metrics are different. The Gauss curvature $K$ is negative provided
\begin{equation}
   \dot{a}\ddot{a}>-b^2\tau^{-3} \, . \label{-a'}
 \end{equation}

The surface (\ref{s2}) is spacelike iff
\be
\dot{a}^2<-1+b^2\tau^{-2}\ .\label{cond1}
\ee
Since $b\neq 0$ one can assume
\begin{equation}
b>0\label{b1}
\end{equation}
without a loss of generality.
The function $\rho$ is given by
\begin{align}
   \rho&=\frac{b^2-\tau^2(1+\dot{a}^2)}{\sqrt{b^2+\tau^3\dot{a}\ddot{a}}} \, .   \label{rho2}
\end{align}
 The corresponding spacetime metric (\ref{metric}) can be transformed to the form
 \begin{align}
g=\gamma e^{\psi}\Big[\big(\frac{b^2}{4\tau^4\dot{a}^2}+\frac{\ddot{a}}{4\tau\dot{a}}\big)d\tau^2-
     dz^2\Big]-\sqrt{\frac{b^2-\tau^2(1+\dot{a}^2)}{b^2+\tau^3\dot{a}
\ddot{a}}}\big[(-\tau+\tau\dot{a}
     \sinh \sigma+\nonumber\\
b\cosh \sigma)
(\theta^2)^2+2(\tau\dot{a}\cosh \sigma+b\sinh \sigma)\theta^2 \theta^3+(\tau+\tau
     \dot{a}\sinh \sigma+b\cosh \sigma)(\theta^3)^2\big]\ ,\label{metric2}
 \end{align} 
where
 \begin{align}
   & \sigma(\tau)=b\int \frac{d\tau}{\tau^2\dot{a}}\, , \label{phi}\\
   &\theta^2=\cosh z dx-\sinh z dy\, , \qquad \theta^3=-\sinh z dx+\cosh z dy 
          \label{thetas}
 \end{align} 
and the new coordinate $z$ is related to $\varphi$ via
\begin{equation}
z=\frac{1}{2}(\varphi+\sigma)\, .\label{chi'}
\end{equation}

For the surface (\ref{s3}) the function $a(\tau)$ has to
obey condition (\ref{-a'}) and 
\begin{align}
   & \dot{a}^2>1+b^2\tau^{-2} \, .       \label{2cond2}
 \end{align}
Due to a residual freedom of coordinates $x^i$ one can replace condition (\ref{2cond2}) by
\begin{equation}
\dot{a}>\sqrt{1+b^2\tau^{-2}}\label{a.}
\end{equation}
without a loss of generality.
In this case 
\begin{equation}
   \rho=\frac{\tau^2(\dot{a}^2-1)-b^2}{\sqrt{b^2+\tau^3\dot{a}\ddot{a}}}   \label{rho3}
 \end{equation}
and the corresponding spacetime metric is given by
 \begin{align}
g=\gamma e^{\psi}\Big[\big(\frac{b^2}{4\tau^4\dot{a}^2}+\frac{\ddot{a}}{4\tau\dot{a}}\big)d\tau^2-
     dz^2\Big]-\sqrt{\frac{\tau^2(\dot{a}^2-1)-b^2}{b^2+\tau^3\dot{a}
\ddot{a}}}\big[(-\tau+\tau\dot{a}
     \cosh \sigma+\nonumber\\
b\sinh \sigma)
   (\theta^2)^2+2(\tau\dot{a}\sinh \sigma+b\cosh \sigma)\theta^2 \theta^3+(\tau+\tau
     \dot{a}\cosh \sigma+b\sinh \sigma)(\theta^3)^2\big]\label{metric3}
 \end{align}     
and relations (\ref{phi})-(\ref{chi'}).

In both cases (\ref{metric2}) and (\ref{metric3}) the function $\psi$ has to satisfy equation (\ref{detR=0}) and condition (\ref{cond}).  If $\gamma=1$ and $\psi=\psi(\tau)$  these metrics  are cosmological solutions of Bianchi
type $VI_0$ (tilted if $b\neq 0$).

If $\gamma=1$, $\psi=\psi(\tau)$ and $b=0$ the metric (\ref{metric3}) takes the form
 \begin{equation}
  \begin{split}
   g=&\rho e^{\omega}\big( \frac{\ddot{a}}{4\tau\dot{a}}d\tau^2-dz^2\big)-
     \rho\sqrt{\frac{\dot{a}-1}{\dot{a}+1}}\big( \cosh z dx-\sinh z dy\big)^2+\\
     &-\rho\sqrt{\frac{\dot{a}+1}{\dot{a}-1}}\big( -\sinh z dx+\cosh z dy\big)^2
      \, ,\label{metric7}
  \end{split}
 \end{equation}
where
\begin{align}
   \rho&=(\dot{a}^2-1)\sqrt{\frac{\tau}{\dot{a}\ddot{a}}} 
 \end{align}
and the function $a$ is assumed to satisfy
$\dot {a}>1$, $\ddot{a}>0$ (see (\ref{a.}) and (\ref{-a'})).
In this case the general solution of (\ref{detR=0}) is given by (\ref{omega}) and the fluid energy density and pressure
are given by (\ref{e1})-(\ref{f'}).

\section{Metrics related to surfaces of  type N }
In order to parametrize the surface (\ref{s6}) it is convenient to use coordinates $\tau$ and $z$ given by
\begin{align}
\tau&=u+\frac{vx_2}{b}-\frac{v^3}{6b^2}\ ,\label{tau}\\
z&=v+\sigma \, , \qquad \sigma(\tau)=\frac{1}{2}\int \dot{a}^2d\tau\ . \label{sigma_c3}\end{align}
The surface is spacelike and $K<0$ iff
\begin{equation}
   b^{-1}a\dot{a}^2>2\ ,\quad \dot{a}^4+4b\ddot{a}>0 \, .\label{cond5} 
 \end{equation}  
Under conditions (\ref{cond5}) definition (\ref{K1}) yields
\begin{align} 
   \rho&=\frac{\abs{a\dot{a}^2-2b}}{\sqrt{\dot{a}^4+4b\ddot{a}}} \, .    \label{rho_c3}\end{align}
It follows from (\ref{cond5}) that $\dot{a}\neq 0$. Hence, one can assume
\begin{equation}
\dot{a}>0
\end{equation}
without a loss of generality.
Let $2bx,y$ be the coordinates $x^a$ in the expression (\ref{metric}) for the spacetime metric. In the coordinates $\tau$, $x$, $y$, $z$ this metric  reads
 \begin{align}
g=&\gamma e^{\psi}\Big[(\frac{\dot{a}^4}{4}+b\ddot{a})d\tau^2-d z^2\Big]
    -\rho\sqrt{\frac{2}{b^{-1}a\dot{a}^2-2}} \Big[\dot{a}\big(dy-z d x\big)^2+\nonumber\\
    &+\big(2\dot{a}\sigma-4b\big)\big(dy-z dx)dx+
     \big(\dot{a}\sigma^2-4b\sigma+2ba\dot{a}\big)dx^2\Big] \, .
  \label{metric5}
 \end{align}   
%where  
% \begin{align} 
%      \theta^2&=zdx-dy\, , \qquad \theta^3=dx \, .
% \end{align} 
The  Einstein equations reduce to equation (\ref{detR=0}) and condition (\ref{cond}). If $\gamma=1$ and $\psi=\psi(\tau)$ the metric (\ref{metric5}) is a tilted cosmological solution of Bianchi type $II$. 

The surface (\ref{s5}) is preserved by the Killing vector field (\ref{k3}) with $b=0$. It is convenient to parametrize $\Sigma$ in the following way
 \begin{equation}
   x^i(\tau,z)=\frac{1}{2}\Big((1-z^2)\tau+a(\tau),\,2\tau z,\, (1+z^2)\tau-a(\tau)\Big) \, , \qquad 
     \tau\geq 0 \, . 
 \end{equation}    
The surface is spacelike and $K<0$ iff
 \begin{equation}
   \dot{a}>0 \, ,\qquad \ddot{a}>0 \, .\label{a..}
 \end{equation}
In this case
\begin{equation}
  \rho=\dot{a}\sqrt{\frac{2\tau}{\ddot{a}}} \label{rho_c2}
 \end{equation}
and the corresponding spacetime metric reads
 \begin{equation}
   g=\gamma e^{\psi}\big(\frac{\ddot{a}}{2\tau}d\tau^2-dz^2\big)-\dot{a}\sqrt{\frac{2\tau}{\ddot{a}}}\big[\sqrt{\dot{a}}dx^2+
      \frac{1}{\sqrt{\dot{a}}}(dy-zdx)^2\big] \,  .\label{metric4}
 \end{equation} 
The function $\psi$ has to satisfy equation (6) and condition (\ref{cond}). 

If $\gamma=1$ and $\psi=\psi(\tau)$ the metric (\ref{metric4}) is a non-tilted cosmological solution of Bianchi type $II$.
It reads
 \begin{equation}
   g=\rho\sqrt{\dot{a}}e^{\omega}\big(\frac{\ddot{a}}{2\tau}d\tau^2-dz^2\big)-\sqrt{\frac{2\tau\dot{a}}{\ddot{a}}}\big[\dot{a}dx^2+(dy-zdx)^2\big]\ ,\label{metric4a}
 \end{equation} 
where
 \begin{equation}
    \omega(\tau)=\omega_0\int \frac{\ddot{a}}{\tau\dot{a}}\,, \quad \omega_0=\text{const}\, .   \label{omega_c2} 
 \end{equation}
The corresponding energy density and pressure of the fluid can be expressed in the following way
 \begin{align}
   \epsilon(\tau)&=\frac{e^{-\omega}}{\rho^2\sqrt{\dot{a}}}\Big(\frac{3\rho'\,^2}{4\rho}+\frac{\rho'}
                   {2\rho}(\tau+2\omega_0)-\frac{\tau'}{4}-\frac{\tau^2}{4\rho}\Big) \, , \\
   p(\tau)&=\epsilon-\frac{\rho''}{\sqrt{\dot{a}}\rho^2}e^{-\omega} \, ,
 \end{align}
where we denote 
\begin{equation}
f'=\dot f \sqrt{\frac{2\tau}{\ddot{a}}}\label{f''}
\end{equation}
for any function $f(\tau)$. Metrics (\ref{metric4a}) include particular solutions of Collins and Stewart \cite{CS} ( metrics (14.20) in \cite{S}) which satisfy $p=(\gamma -1)\epsilon$ with a constant $\gamma$. They follow from 
(\ref{metric4a}) for $a=\frac{1}{n}\tau^n$ where $n=\frac{2+\gamma}{4-2\gamma}$ and  $\frac{2}{3}<\gamma<2$.

The surface (\ref{s4}) induces  the  following spacetime metric
 \begin{equation}
   g=2\gamma e^{\psi}d\tau dz-\tau^3 dx^2-\tau (dy-zdx)^2 \label{metric6}
 \end{equation}
in some coordinates $\tau$, $x$, $y$, $z$.
In this case the function $\psi$ is given   by
 \begin{equation}
  \psi(\tau,z)=\frac{1}{4}\log \tau-\int \frac{\big( z+c(\tau)\big)^2}{4\tau^3} d\tau 
 \end{equation}
where $c(\tau)$ is an arbitrary function.  Metric (\ref{metric6}) is not interesting because it yields $p=\epsilon<0$.

\section{Summary}
Following the method elaborated in \cite{ST} we have constructed several classes of perfect fluid solutions of the Einstein equations. The solutions are related to spacelike surfaces with a negative curvature in the 3-dimensional Minkowski space $M^{2,1}$. We classified those surfaces which are invariant under a 1-dimensional group of isometries of $M^{2,1}$ (equations (\ref{s1})-(\ref{s4})). For all of them we calculated the corresponding  spacetime metrics (equations (\ref{metric1}),  (\ref{metric2}), (\ref{metric3}), (\ref{metric5}), (\ref{metric4}), (\ref{metric6})). In all cases except (\ref{metric6}) they depend on two functions $a$, $\psi$ and the Einstein equations reduce to equation (\ref{detR=0}) and inequality (\ref{cond}). Under assumption (\ref{psi_t}) these metrics  become cosmological solutions of Bianchi type $II$, $VI_0$ or $VII_0$ and a barotropic equation of state can be imposed. In some cases (see (\ref{sol1}),  (\ref{metric7}) and (\ref{metric4a})) we succeded to solve all of the Einstein equations up to quadratures. The resulting  metrics depend on an arbitrary function of time, which can be further specified in order to satisfy an equation of state. 

\bigskip
\bigskip

\noindent {\bf Acknowledgments}. This work was partially supported by
the Polish  Committee for Scientific Research (grant 2 P03B 127 24 and 2 P03B 036 23).

\end{document}